\documentclass[10pt,aps,prd,floatfix,groupedaddress,nofootinbib,superscriptaddress,twocolumn,notitlepage,showkeys]{revtex4-1}

\usepackage{amsmath,amssymb,bm,color,float,graphicx,mathrsfs}
\usepackage[hidelinks]{hyperref}
\hypersetup{colorlinks=true,linkcolor=blue,citecolor=blue,urlcolor=blue}
\usepackage{Macro}

\usepackage[normalem]{ulem}

\begin{document}

\title{Planck Constraints on Axion-Like Particles through Isotropic Cosmic Birefringence}

\author{Toshiya Namikawa}
\affiliation{Center for Data-Driven Discovery, Kavli IPMU (WPI), UTIAS, The University of Tokyo, Kashiwa, 277-8583, Japan}
\affiliation{Department of Applied Mathematics and Theoretical Physics, University of Cambridge, Wilberforce Road, Cambridge CB3 0WA, United Kingdom}
\affiliation{Kavli Institute for Cosmology, University of Cambridge, Madingley Road, Cambridge CB3 OHA, United Kingdom}

\author{Kai Murai}%
\affiliation{Department of Physics, Tohoku University, Sendai, Miyagi 980-8578, Japan}%

\author{Fumihiro Naokawa}
\affiliation{Research Center for the Early Universe, The University of Tokyo, Bunkyo-ku, Tokyo 113-0033, Japan}
\affiliation{Department of Physics, Graduate School of Science, The University of Tokyo, Bunkyo-ku, Tokyo 113-0033, Japan}

\date{\today}

\begin{abstract}
We present constraints on isotropic cosmic birefringence induced by axion-like particles (ALPs), derived from the analysis of cosmic microwave background (CMB) polarization measurements obtained with the high-frequency channels of Planck. Recent measurements report a hint of isotropic cosmic birefringence, though its origin remains uncertain. The detailed dynamics of ALPs can leave characteristic imprints on the shape of the $EB$ angular power spectrum, which can be exploited to constrain specific models of cosmic birefringence. We first construct a multifrequency likelihood that incorporates an intrinsic nonzero $EB$ power spectrum. We also show that the likelihood used in previous studies can be further simplified without loss of generality. Using this framework, we simultaneously constrain the ALP model parameters, the instrumental miscalibration angle, and the amplitudes of the $EB$ power spectrum of a Galactic dust foreground model. We find that, if ALPs are responsible for the observed cosmic birefringence, ALP masses at $\log_{10}m_{\phi}[{\rm eV}]\simeq-27.8$, $-27.5$, $-27.3$, $-27.2$, $-27.1$, as well as $\log_{10}m_{\phi}[{\rm eV}]\in[-27.0,-26.5]$, are excluded at more than $2\,\sigma$ statistical significance. 
\end{abstract} 

\keywords{cosmology, cosmic microwave background}


\maketitle

\section{Introduction} \label{sec:intro}

Recent analyses of cosmic microwave background (CMB) data have revealed a tantalizing hint of cosmic birefringence---a rotation of the polarization plane of photons as they propagate through space~\cite{Minami:2020:biref,Diego-Palazuelos:2022,Eskilt:2022:biref-freq,Eskilt:2022:biref-const,Eskilt:2023:EDE} (see Ref.~\cite{Komatsu:2022:review} for a comprehensive review). As a parity-violating effect, cosmic birefringence offers a potential smoking gun for new physics beyond both the $\Lambda$ Cold Dark Matter ($\Lambda$CDM) model and the Standard Model of particle physics~\cite{Nakai:2023}.

Cosmic birefringence can be induced by a pseudoscalar field, such as axion-like particles (ALPs), coupled to the electromagnetic field through the Chern-Simons interaction:
\al{
    \mathcal{L}\supset -\frac{1}{4}g\phi F_{\mu\nu}\tilde{F}^{\mu\nu}
    \,,
} 
where $g$ is the coupling constant, $\phi$ is an ALP field, $F_{\mu\nu}$ denotes the electromagnetic field tensor, and $\tilde{F}^{\mu\nu}$ is its dual. Numerous studies have investigated this effect in various cosmological contexts, including ALP fields associated with dark energy \cite{Carroll:1998:DE,Liu:2006:biref-time-evolve,Panda:2010,Fujita:2020aqt,Fujita:2020ecn,Choi:2021aze,Obata:2021,Gasparotto:2022uqo,Galaverni:2023}, early dark energy scenarios \cite{Fujita:2020ecn,Murai:2022:EDE,Eskilt:2023:EDE,Kochappan:2024:biref}, and axion dark matter \cite{Finelli:2009,Sigl:2018:biref-sup,Liu:2016:AxionDM,Fedderke:2019:biref,Zhang:2024dmi}. Additional mechanisms include topological defects \cite{Takahashi:2020tqv,Kitajima:2022jzz,Jain:2022jrp,Gonzalez:2022mcx,Lee:2025:biref-DW} and possible imprints of quantum gravity \cite{Myers:2003fd,Balaji:2003sw,Arvanitaki:2009fg}. 
Looking ahead, ongoing and upcoming CMB experiments, including BICEP \cite{Cornelison:2022:BICEP3,BICEPArray}, the Simons Observatory \cite{SimonsObservatory}, CMB-S4 \cite{CMBS4}, and LiteBIRD \cite{LiteBIRD,LiteBIRD:2025:biref}, are expected to substantially reduce polarization noise and improve sensitivity to birefringence-induced signals, thereby enabling more stringent tests of these theoretical scenarios.

To investigate the origin of cosmic birefringence, it is useful to consider observables that are unaffected by a miscalibrated polarization angle. One such observable is the anisotropic polarization rotation induced by the fluctuations of the ALP field, $\delta\phi$ \cite{Carroll:1998:DE,Lue:1999:biref-EB,Caldwell:2011,Lee:2015,Leon:2017,Yin:2023:biref,Ferreira:2023,Greco:2022:aniso-biref-tomography,Greco:2022:cross,Greco:2024,Arcari:2024nhw}. Several studies have constrained these anisotropies by reconstructing the polarization rotation angle \cite{Gluscevic:2012,PB15:rot,BICEP2:2017lpa,Contreras:2017,Namikawa:2020:biref,SPT:2020:biref,Gruppuso:2020,Bortolami:2022whx,BK-LoS:2023,Zagatti:2024}, while others have derived limits using CMB polarization power spectra \cite{Li:2014,Alighieri:2014yoa,Liu:2016:AxionDM,Zhang:2024dmi}. However, the time evolution of the ALP field can substantially suppress the induced $B$-mode power spectrum \cite{Namikawa:2024:BB}.

In this work, we aim to disentangle the effects of a miscalibrated polarization angle from genuine birefringence by exploiting the spectral shape of the $EB$ power spectrum. The $EB$ power spectrum is particularly sensitive to the time evolution of pseudoscalar fields during the epochs of recombination and reionization, which can significantly alter the CMB polarization signals \cite{Finelli:2009,Lee:2013:biref,Gubitosi:2014:biref-time,Sherwin:2021:biref,Nakatsuka:2022,Naokawa:2023,Yin:2023:biref,Naokawa:2024xhn,Murai:2024yul}. By analyzing this spectral shape, we place constraints on the ALP mass and other model parameters.
Additional constraints on late-time ALP dynamics can be obtained through tomographic probes, such as the polarized Sunyaev–Zel'dovich effect \cite{Lee:2022:pSZ-biref,Namikawa:2023:pSZ} and galaxy polarization measurements \cite{Carroll:1997:radio,Yin:2024:galaxy}, both of which offer complementary information and help mitigate degeneracies with polarization calibration errors.

This paper is organized as follows. In Sec.~\ref{sec:theory}, we review the theoretical framework of the $EB$ power spectrum induced by ALP-driven cosmic birefringence. Section~\ref{sec:analysis} describes the datasets used and our analysis methodology. We present our constraints on ALP model parameters in Sec.~\ref{sec:results}. Finally, we summarize our findings and discuss their implications in Sec.~\ref{sec:summary}.

\section{Isotropic cosmic birefringence from ALP} \label{sec:theory}

In this section, we briefly review prior studies on the angular 
$EB$ power spectrum induced by cosmic birefringence from ALPs~\cite{Liu:2006:biref-time-evolve,Finelli:2009,Gubitosi:2014:biref-time,Lee:2013:biref,Sherwin:2021:biref,Nakatsuka:2022,Murai:2022:EDE,Naokawa:2023,Murai:2024yul,Naokawa:2024xhn}.

We begin by considering the case where cosmic birefringence rotates the polarization plane of CMB photons by a constant angle $\beta$. In this scenario, the observed Stokes parameters are transformed as
\al{
    Q\pm\iu U=[Q^{\rm lss}\pm\iu U^{\rm lss}]\exp(\pm2\iu\beta)
    \,, 
}
where $Q^{\rm lss}$ and $U^{\rm lss}$ denote the Stokes parameters at the last scattering surface, in the absence of rotation.
The $E$- and $B$-mode coefficients are defined from the spin-weighted spherical harmonic decomposition of the polarization field \cite{Zaldarriaga:1996xe,Kamionkowski:1996:eb}:
\al{
    E_{\l m} \pm\iu B_{\l m} = - \Int{2}{\hatn}{} (Y_{\l m}^{\pm2}(\hatn))^* P^\pm(\hatn)
    \,, \label{Eq:EB-def}
}
where $P^\pm=Q\pm\iu U$ and $Y_{\l m}^{\pm2}$ are spin-2 spherical harmonics. Under a constant rotation $\beta$, the $E$- and $B$-modes are rotated according to
\al{
    \begin{pmatrix} E_{\l m} \\ B_{\l m} \end{pmatrix} 
    = \bR{R}(\beta) \begin{pmatrix} E^{\rm lss}_{\l m} \\ B^{\rm lss}_{\l m} \end{pmatrix} 
    \,, 
}
where the rotation matrix is defined as
\al{
    \bR{R}(\beta) = \begin{pmatrix} \cos2\beta & -\sin2\beta \\ \sin2\beta & \cos2\beta \end{pmatrix}
    \,. \label{Eq:rotmat}
}
This leads to a nonzero $EB$ power spectrum given by
\al{
    C_\l^{EB} = \frac{\sin4\beta}{2}(C_\l^{EE,{\rm lss}}-C_\l^{BB,{\rm lss}})
    \,. \label{Eq:ClEB:beta-const}
}

When cosmic birefringence is sourced by an ALP field, the rotation angle becomes time dependent. The total rotation angle for a photon observed today, emitted at conformal time $\eta$, is
\al{
    \beta(\eta) 
    &= \frac{g}{2}[\phi(\eta_0)-\phi(\eta)] 
    \\
    &= \frac{g\phi_{\rm ini}}{2}\frac{\phi(\eta_0)-\phi(\eta)}{\phi_{\rm ini}}
    \equiv \beta_{\rm ini}[f(\eta)-f(\eta_0)]
    \,, 
}
where $\eta_0$ is the present conformal time, $\phi_{\rm ini}$ is an initial value of the ALP field, $\beta_{\rm ini}=-g\phi_{\rm ini}/2$, and $f=\phi/\phi_{\rm ini}$. 
To compute the impact of this time-dependent rotation on the CMB polarization, we solve the Boltzmann equation for the polarized photon distribution \cite{Liu:2006:biref-time-evolve,Finelli:2009,Gubitosi:2014:biref-time,Lee:2013:biref},
\begin{align}
    &_{\pm2}\Delta'_P + \iu q\mu\,_{\pm2}\Delta_P 
    \notag \\
    &\qquad = a n_{\rm e}\sigma_T
        \left[
            -\,_{\pm2}\Delta_P + \sqrt{\frac{6\pi}{5}}\,_{\pm2}Y_{20}(\mu)\Pi(\eta,q)
        \right]
    \notag \\
    &\qquad\qquad \pm \iu g\phi'\,_{\pm2}\Delta_P
    \,, \label{Eq:Boltzmann}
\end{align}
where $_{\pm2}\Delta_P$ are the Fourier modes of $Q\pm \iu U$ and are the functions of conformal time, the magnitude of the Fourier wavevector, $q$, and the cosine of the angle between the Fourier wavevector and line-of-sight direction, $\mu$. We also introduce the scale factor, $a$, the electron number density, $n_{\rm e}$, the cross-section of the Thomson scattering, $\sigma_T$, and the polarization source term, $\Pi$, introduced in Ref.~\cite{Zaldarriaga:1996xe}. A prime denotes a derivative with respect to conformal time. 
The evolution of the ALP field is governed by
\begin{equation}
    \phi'' + 2\frac{a'}{a}\phi' + a^2m_\phi^2 \phi = 0
    \,, \label{Eq:phi-EoM}
\end{equation}
assuming a quadratic potential $V(\phi)=m_\phi^2\phi^2/2$. 
While ALPs generally possess periodic potentials such as a cosine-type one, we employ a quadratic one for simplicity of the analysis. If the ALP evolves around the potential minimum where the potential can be approximated by a quadratic one, our analysis can be applied.

To derive the angular power spectra, we expand $_{\pm2}\Delta_P$ as \cite{Zaldarriaga:1996xe}
\begin{align}
    _{\pm2}\Delta_P(\eta_0,q,\mu) 
    &\equiv -\sum_\l \iu^{-\l}\sqrt{4\pi(2\l+1)} 
    \notag \notag \\
    &\quad\times [\Delta_{E,\l}(q)\pm \iu\Delta_{B,\l}(q)] {}_{\pm2}Y_{\l0}(\mu) 
    \,. 
\end{align}
The $EB$ angular power spectrum from these $E$- and $B$-modes is then given by
\begin{equation}
    C_\l^{EB} = 4\pi
    \Int{}{(\ln q)}{} \mathcal{P}_s(q)\Delta_{E,\l}(q)\Delta_{B,\l}(q)
    \,, \label{Eq:ClXY}
\end{equation}
where $\mathcal{P}_s(q)$ is the dimensionless power spectrum of the primordial scalar curvature perturbations. 
Solving Eq.~\eqref{Eq:Boltzmann} provides the full shape of the birefringence-induced $EB$ power spectrum as described in Eq.~\eqref{Eq:ClXY}.

Note that the trajectories of CMB photons are deflected by gravitational lensing due to foreground large-scale structures (see, e.g., Ref.~\cite{Lewis:2006:review}). This lensing effect leads to a smearing of the acoustic peaks in the observed CMB anisotropies and enhances the amplitude of small-scale anisotropies. Reference~\cite{Naokawa:2023} derives the lensed $EB$ power spectrum by utilizing the fact that gravitational lensing and a global rotation of the polarization plane commute---that is, lensing does not affect the rotation, and vice versa \cite{Namikawa:2021:mode}.

If the ALP mass satisfies $m_{\phi}\alt 10^{-28}$\,eV, the field evolves slowly and the rotation angle during recombination remains nearly constant. In this regime, the $EB$ power spectrum is well approximated by Eq.~\eqref{Eq:ClEB:beta-const}, using the rotation angle at recombination. An exception arises at low multipoles where reionization effects become significant \cite{Sherwin:2021:biref}. 
In contrast, for $m_{\phi}\agt 10^{-28}$\,eV, the ALP field begins oscillating before or during recombination, causing the rotation angle to decrease significantly by the time of photon decoupling. As a result, the conversion of $E$- to $B$-modes is suppressed, leading to a diminished $EB$ correlation.

The rotation angle at any epoch satisfies $|\beta(\eta)|\lesssim|\beta_{\rm ini}|$. 
When $|\beta_{\rm ini}|<1\,$deg, the $EB$ power spectrum amplitude scales linearly with $\beta_{\rm ini}$ at approximately $0.08\%$ accuracy, allowing for simple rescaling in the small-angle limit. Even when $|\beta_{\rm ini}|\gg1\,$deg, this linear rescaling of the $EB$ power spectrum by $\beta_{\rm ini}$ remains valid, provided that the rotation angle satisfies $|\beta(\eta)|<1\,$deg during recombination.

\section{Analysis} \label{sec:analysis}

In this section, we present our methodology for jointly constraining the parameters of ALPs, the polarization miscalibration angle, and foreground contributions. We begin by outlining the key equations used to relate theoretical predictions to observational data, along with the likelihood function employed in the analysis. A detailed derivation of these equations is provided in Appendix~\ref{app:eqs}.
We then describe the datasets used in our analysis and specify the set of model parameters we aim to constrain.

\subsection{Basic equations for analysis}

We here derive the core equations used to analyze the observed CMB polarization power spectra and constrain parameters. 
We model the observed spherical harmonic coefficients of the CMB polarization maps as \cite{Minami:2020:method,Diego-Palazuelos:2022,Eskilt:2022:biref-const}
\al{
    \begin{pmatrix} \hat{E}_{\l m,i} \\ \hat{B}_{\l m,i} \end{pmatrix} 
    &= \bR{R}(\alpha_i) \begin{pmatrix} E_{\l m} + f^E_{\l m,i}\\ B_{\l m}+f^B_{\l m,i} \end{pmatrix} 
    + \begin{pmatrix} n^E_{\l m,i} \\ n^B_{\l m,i} \end{pmatrix}
    \,, \label{Eq:EBij}
}
where $\bR{R}(\alpha_i)$ is the rotation matrix defined in Eq.~\eqref{Eq:rotmat} and,
\begin{itemize}
    \item $\hat{E}_{\l m,i}$, $\hat{B}_{\l m,i}$: observed $E$- and $B$-mode components of the $i$th map, 
    \item $\alpha_i$: miscalibration angle for the $i$th map,
    \item $E_{\l m}$, $B_{\l m}$: cosmological $E$- and $B$-mode signals that could already be rotated by cosmic birefringence, 
    \item $f^E_{\l m,i}$, $f^B_{\l m,i}$: $E$- and $B$-mode foreground contributions in the $i$th map,
    \item $n^E_{\l m,i}$, $n^B_{\l m,i}$: $E$- and $B$-mode instrumental noise in the $i$th map. 
\end{itemize}
To express the angular power spectra, we introduce the following matrices and vectors \cite{Minami:2020:method,Diego-Palazuelos:2022,Eskilt:2022:biref-const}:
\al{
    \bR{R}(\alpha_i,\alpha_j) 
    &\equiv 
    \begin{pmatrix}
        \cos2\alpha_i\cos2\alpha_j & \sin2\alpha_i\sin2\alpha_j \\
        \sin2\alpha_i\sin2\alpha_j & \cos2\alpha_i\cos2\alpha_j 
    \end{pmatrix}
    \,,
    \\ 
    \vec{R}(\alpha_i,\alpha_j) 
    &\equiv 
    \begin{pmatrix}
        \cos2\alpha_i\sin2\alpha_j \\ -\sin2\alpha_i\cos2\alpha_j
    \end{pmatrix}
    \,,
    \\
    \bR{D}(\alpha_i,\alpha_j) 
    &\equiv 
    \begin{pmatrix}
        -\cos2\alpha_i\sin2\alpha_j & -\sin2\alpha_i\cos2\alpha_j \\
        \sin2\alpha_i\cos2\alpha_j & \cos2\alpha_i\sin2\alpha_j
    \end{pmatrix}
    \,,
    \\ 
    \vec{D}(\alpha_i,\alpha_j) 
    &\equiv 
    \begin{pmatrix}
        \cos2\alpha_i\cos2\alpha_j \\ -\sin2\alpha_i\sin2\alpha_j
    \end{pmatrix}
    \,.
}
Using the above definitions, the data vector composed of the power spectra is written as (see Appendix~\ref{app:eqs} for derivation)
\al{
    \vec{d}_{\l,ij} 
    &\equiv \begin{pmatrix}
        \hat{C}_\l^{E_iE_j} \\ \hat{C}_\l^{B_iB_j} \\ \hat{C}_\l^{E_iB_j} 
    \end{pmatrix}
    \notag \\
    &= \begin{pmatrix} \bR{R}(\alpha_i,\alpha_j) \\ \vec{R}^T(\alpha_i,\alpha_j) \end{pmatrix}
    \begin{pmatrix}
        C_\l^{EE}+F_\l^{E_iE_j} \\ 
        C_\l^{BB}+F_\l^{B_iB_j}
    \end{pmatrix}
    \notag \\
    &\qquad+ \begin{pmatrix} \bR{D}(\alpha_i,\alpha_j) \\ \vec{D}^T(\alpha_i,\alpha_j) \end{pmatrix}
    \begin{pmatrix}
        F_\l^{E_iB_j} \\
        F_\l^{B_iE_j}
    \end{pmatrix}
    \notag \\
    &\qquad+ \vec{E}(\alpha_i,\alpha_j) C_\l^{EB}
    \,, \label{Eq:vecC:general}
}
where $F_\l^{XY}$ denotes the foreground power spectrum, and we define
\al{
    \vec{E}(\alpha_i,\alpha_j) \equiv \begin{pmatrix} -\sin2\theta_{ij} \\ \sin2\theta_{ij} \\ \cos2\theta_{ij} \end{pmatrix}
    \,, 
}
with $\theta_{ij}=\alpha_i+\alpha_j$. 
We assume that the noise components from different frequency channels are statistically independent and neglect noise covariance \cite{Minami:2020:method,Diego-Palazuelos:2022,Eskilt:2022:biref-const}.

\subsubsection{General case}

Eliminating the $E$- and $B$-mode auto power spectra, $C_\l^{EE}+F_\l^{E_iE_j}$ and $C_\l^{BB}+F_\l^{B_iB_j}$, from \eq{Eq:vecC:general}, we obtain (see Appendix~\ref{app:eqs} for derivation)
\al{
    \hat{C}_\l^{E_iB_j} 
    &= \frac{\hat{C}_\l^{E_iE_j}\sin4\alpha_j-\hat{C}_\l^{B_iB_j}\sin4\alpha_i}{\cos4\alpha_i+\cos4\alpha_j}
    \notag \\
    &+ 2\frac{F_\l^{E_iB_j}\cos2\alpha_i\cos2\alpha_j+F_\l^{B_iE_j}\sin2\alpha_i\sin2\alpha_j}{\cos4\alpha_i+\cos4\alpha_j}
    \notag \\
    &+ \frac{C_\l^{EB}}{\cos2\theta_{ij}} 
    \,. \label{Eq:dij:general}
}
If intrinsic foreground $EB$ correlations are negligible, this simplifies to
\al{
    \hat{C}_\l^{E_iB_j} 
    &= \frac{\hat{C}_\l^{E_iE_j}\sin4\alpha_j-\hat{C}_\l^{B_iB_j}\sin4\alpha_i}{\cos4\alpha_i+\cos4\alpha_j}
    + \frac{C_\l^{EB}}{\cos2\theta_{ij}} 
    \,. \label{Eq:dij:general:noFEB}
}
Moreover, subtracting the symmetric component under $i\leftrightarrow j$ yields an expression involving only observed quantities,
\al{
    \hat{C}_\l^{E_iB_j} - \hat{C}_\l^{E_jB_i} 
    &= \frac{\sin4\alpha_j-\sin4\alpha_i}{\cos4\alpha_i+\cos4\alpha_j}(\hat{C}_\l^{E_iE_j}+\hat{C}_\l^{B_iB_j})
    \,. 
}

\subsubsection{Modeling the intrinsic \texorpdfstring{$EB$}{EB} foregrounds}

To account for Galactic foregrounds---primarily thermal dust at Planck high-frequency channels---we adopt the empirical model of Ref.~\cite{Eskilt:2022:biref-const}, where the foreground $EB$ power spectrum is modeled as
\al{
    F_\l^{E_iB_j} = A^{\rm dust}_\l\sin4\psi_\l F_\l^{E_iE_j}
    \,, \label{Eq:dust-EB-model}
}
with $A^{\rm dust}_\l$ as a free amplitude parameter, and the effective dust polarization angle defined by
\al{
    \psi_\l \equiv \frac{1}{2}\arctan\left(\frac{F_\l^{TB}}{F_\l^{TE}}\right)
    \,, 
}
determined from Planck 353 GHz maps. As in previous studies, we assume $A^{\rm dust}_\l$ and $\psi_\l$ are frequency independent. This simplification has been shown to be valid given the weak frequency dependence of the prefactor \cite{Diego-Palazuelos:2022}.

The complete data model, including this dust-induced correlation, becomes
\al{
    \vec{d}_{\l,ij}
    &= 
    \begin{pmatrix} \bR{R} \\ \vec{R}^T \end{pmatrix}
    \begin{pmatrix}
        C_\l^{EE} \\ 
        C_\l^{BB}
    \end{pmatrix}
    + \begin{pmatrix} \bR{\Lambda} \\ \vec{\Lambda}^T \end{pmatrix}
    \begin{pmatrix}
        F_\l^{E_iE_j} \\ 
        F_\l^{B_iB_j}
    \end{pmatrix}
    + \vec{E}C_\l^{EB}
    \,, \label{Eq:vecC:psil}
}
where the matrix $\bR{\Lambda}$ includes the dust angle contribution: 
\al{
    \begin{pmatrix} \bR{\Lambda} \\ \vec{\Lambda}^T \end{pmatrix}
    \equiv 
    \begin{pmatrix} \bR{R} \\ \vec{R}^T \end{pmatrix}
    +  \tan 2x_\l \begin{pmatrix}  & 0 \\ \vec{E} & 0 \\ & 0 \end{pmatrix}
    \,, \label{Eq:def:Lambda}
}
with $\tan 2x_\l\equiv A^{\rm dust}_\l\sin 4\psi_\l$. 
As \eq{Eq:vecC:psil} contains three equations, we can eliminate the foreground power spectra, $F_\l^{E_iE_j}$ and $F_\l^{B_iB_j}$, leading to a single equation for the $EB$ power spectrum (see Appendix~\ref{app:eqs} for derivation): 
\al{
    \hat{C}_\l^{E_iB_j} 
    &= \frac{\hat{C}_\l^{E_iE_j}\cos2\alpha_j\sin2\tilde{\theta}_{j,\l}-\hat{C}_\l^{B_iB_j}\sin2\alpha_i\cos2\tilde{\theta}_{i,\l}}{\cos2\tilde{\theta}_{ij,\l}\cos2\delta_{ij}}
    \notag \\
    &\qquad+ \frac{C_\l^{EB}\cos2x_\l-C_\l^{EE}\sin2x_\l}{\cos2\tilde{\theta}_{ij,\l}}
    \,, \label{Eq:chisq}
}    
where we define 
\al{
    \delta_{ij} &= \alpha_i-\alpha_j 
    \,, \\ 
    \tilde{\theta}_{i,\l} &= \alpha_i+x_\l 
    \,, \\
    \tilde{\theta}_{ij,\l} &= \theta_{ij}+x_\l 
    \,.
}
This equation can be recast in a linear form for parameter estimation:
\al{
    \vec{A}^T_{\l,ij}\begin{pmatrix}
        \hat{C}_\l^{E_iE_j} \\ \hat{C}_\l^{B_iB_j} \\ \hat{C}_\l^{E_iB_j} 
    \end{pmatrix} - \vec{B}^T_{\l,ij}\begin{pmatrix}
        C_{\l}^{EE} \\ C_{\l}^{BB} \\C_{\l}^{EB}
    \end{pmatrix} = 0
    \,, \label{Eq:chisq:2}
}
with the vectors defined as 
\al{
    \vec{A}_{\l,ij} &= 
    \begin{pmatrix} 
        -\cos2\alpha_j\sin2\tilde{\theta}_{j,\l}/(\cos2\tilde{\theta}_{ij,\l}\cos2\delta_{ij}) \\
        \sin2\alpha_i\cos2\tilde{\theta}_{i,\l}/(\cos2\tilde{\theta}_{ij,\l}\cos2\delta_{ij}) \\
        1
    \end{pmatrix}
    \,, \label{Eq:vecA:final} \\ 
    \vec{B}_{\l,ij} &= \frac{1}{\cos2\tilde{\theta}_{ij,\l}}
    \begin{pmatrix}
        -\sin2x_\l\\ 0 \\ \cos2x_\l
    \end{pmatrix}
    \,. \label{Eq:vecB:final}
}
This linearized equation is used in our likelihood analysis to simultaneously constrain the birefringence signal, miscalibration angle, and foreground contamination.

\subsubsection{Constant rotation}

Before detailing the likelihood implementation, we consider a special case where cosmic birefringence is modeled as a constant rotation $\beta_i$, potentially varying by frequency band, as done in previous work \cite{Minami:2020:method,Diego-Palazuelos:2022,Eskilt:2022:biref-const}. In this scenario, the total rotation of the CMB signal is described by $\alpha_i+\beta_i$, and the birefringence-induced $EB$ correlation is set to zero. From \eq{Eq:vecC:general}, the data vector becomes 
\al{
    \vec{d}_{\l,ij}
    &= 
    \begin{pmatrix} 
        \bR{R}(\alpha_i+\beta_i,\alpha_j+\beta_j) \\ \vec{R}^T(\alpha_i+\beta_i,\alpha_j+\beta_j) 
    \end{pmatrix}
    \begin{pmatrix}
        C_\l^{EE,{\rm lss}} \\ 
        C_\l^{BB,{\rm lss}}
    \end{pmatrix}
    \notag \\
    &\qquad+ 
    \begin{pmatrix} 
        \bR{R}(\alpha_i,\alpha_j) \\ \vec{R}^T(\alpha_i,\alpha_j) 
    \end{pmatrix}
    \begin{pmatrix}
        F_\l^{E_iE_j} \\ 
        F_\l^{B_iB_j}
    \end{pmatrix}
    \notag \\
    &\qquad+ \begin{pmatrix} \bR{D}(\alpha_i,\alpha_j) \\ \vec{D}^T(\alpha_i,\alpha_j) \end{pmatrix}
    \begin{pmatrix}
        F_\l^{E_iB_j} \\
        F_\l^{B_iE_j}
    \end{pmatrix}
    \,.
}
One can eliminate the $E$- and $B$-mode power spectra of the foregrounds to yield a single equation \cite{Eskilt:2022:biref-const}: 
\al{
    &\hat{C}_\l^{E_iB_j} - \Lambda^T\bR{\Lambda}^{-1} \begin{pmatrix}
        \hat{C}_\l^{E_iE_j} \\ \hat{C}_\l^{B_iB_j} 
    \end{pmatrix}
    \notag \\
    &= [\vec{R}^T(\alpha_i+\beta_i,\alpha_j+\beta_j) 
    \notag \\
    &\qquad- \vec{\Lambda}^T\bR{\Lambda}^{-1} \bR{R}(\alpha_i+\beta_i,\alpha_j+\beta_j)]
    \begin{pmatrix}
        C_\l^{EE,{\rm lss}} \\
        C_\l^{BB,{\rm lss}}
    \end{pmatrix}
    \,. 
}    
Instead of directly using this expression, we simplify by rotating the CMB spectra by $\beta_i$ in \eq{Eq:chisq}, which yields
\al{
    \begin{pmatrix}
        C_\l^{EE} \\
        C_\l^{BB} \\
        C_\l^{EB}
    \end{pmatrix}
    = \begin{pmatrix}
        \bR{R}(\beta_i,\beta_j) \\
        \vec{R}^T(\beta_i,\beta_j) 
    \end{pmatrix}
    \begin{pmatrix}
        C_\l^{EE,{\rm lss}} \\
        C_\l^{BB,{\rm lss}}
    \end{pmatrix}
    \,. 
}
Substituting this into Eq.~\eqref{Eq:chisq:2} yields
\al{
    \vec{A}^T_{\l,ij}\begin{pmatrix}
        \hat{C}_\l^{E_iE_j} \\ \hat{C}_\l^{B_iB_j} \\ \hat{C}_\l^{E_iB_j} 
    \end{pmatrix} - [\vec{B}']^T_{\l,ij}
    \begin{pmatrix}
        C_\l^{EE,{\rm lss}} \\
        C_\l^{BB,{\rm lss}}
    \end{pmatrix} = 0
    \,,
}
with
\al{
    [\vec{B}']^T_{\l,ij} &\equiv \frac{(-\sin2x_\l,0,\cos2x_\l)}{\cos2\tilde{\theta}_{ij,\l}}\begin{pmatrix}
        \bR{R}(\beta_i,\beta_j) \\
        \vec{R}^T(\beta_i,\beta_j) 
    \end{pmatrix}
    \\ 
    &= \frac{1}{\cos2\tilde{\theta}_{ij,\l}}\begin{pmatrix}
        \cos2\beta_i\sin2\Delta_{j,\l} \\ 
        -\sin2\beta_i\cos2\Delta_{j,\l} 
    \end{pmatrix}^T
    \,,
}
where we define $\Delta_{j,\l}=\beta_j-x_\l$. The single equation is then given by
\al{
    \hat{C}_\l^{E_iB_j} 
    &= \frac{\hat{C}_\l^{E_iE_j}\cos2\alpha_j\sin2\tilde{\theta}_{j,\l}-\hat{C}_\l^{B_iB_j}\sin2\alpha_i\cos2\tilde{\theta}_{i,\l}}{\cos2\tilde{\theta}_{ij,\l}\cos2\delta_{ij}}
    \notag \\
    &+ \frac{C_\l^{EE,{\rm lss}}\cos2\beta_i\sin2\Delta_{j,\l}-C_\l^{BB,{\rm lss}}\sin2\beta_i\cos2\Delta_{j,\l}}{\cos2\tilde{\theta}_{ij,\l}}
    \,. \label{Eq:chisq:const}
}
If $x_\l=0$, we obtain the following single equation:
\al{
    &\hat{C}_\l^{E_iB_j} 
    = 
    \frac{\hat{C}_\l^{E_iE_j}\sin4\alpha_j-\hat{C}_\l^{B_iB_j}\sin4\alpha_i}{\cos4\alpha_i+\cos4\alpha_j}
    \notag \\ 
    &\quad+ \frac{C_\l^{EE,{\rm lss}}\cos2\beta_i\sin2\beta_j-C_\l^{BB,{\rm lss}}\sin2\beta_i\cos2\beta_j}{\cos2(\alpha_i+\alpha_j)}
    \,. 
}
The above equation is an alternative simplified expression for Eq.~(6) of Ref.~\cite{Eskilt:2022:biref-const}. 
If $\beta_i=\beta_j=\beta$ and ignore the intrinsic $EB$ correlation of the foregrounds, we find that the equation has the following simple form:
\al{
    \hat{C}_\l^{E_iB_j} 
    &= \frac{\hat{C}_\l^{E_iE_j}\sin4\alpha_j-\hat{C}_\l^{B_iB_j}\sin4\alpha_i}{\cos4\alpha_i+\cos4\alpha_j} 
    \notag \\
    &\quad+ \frac{\sin4\beta}{2\cos2(\alpha_i+\alpha_j)}(C_\l^{EE,{\rm lss}}-C_\l^{BB,{\rm lss}})
    \,. \label{Eq:chisq:const:noFGEB}
}
This equation is an alternative simplified expression for Eq.~(10) of Ref.~\cite{Minami:2020:method}. If we further assume $\alpha_i=\alpha_j=\alpha$, the above equation coincides with Eq.~(3) of Ref.~\cite{Minami:2020:method} but without the intrinsic $EB$ correlations from foregrounds and the CMB: 
\al{
    \hat{C}_\l^{E_iB_j} 
    &= \frac{\tan4\alpha}{2}\left(\hat{C}_\l^{E_iE_j}-\hat{C}_\l^{B_iB_j}\right) 
    \notag \\
    &\qquad+ \frac{\sin4\beta}{2\cos4\alpha}(C_\l^{EE,{\rm lss}}-C_\l^{BB,{\rm lss}})
    \,. \label{Eq:chisq:const:noFGEB:singlefreq}
}

\subsection{Data}

We utilize polarization data of Planck Public Release 4 \cite{Planck:2020:Npipe} measured at the following four frequency channels of the Planck high-frequency instrument (HFI): 100 GHz, 143 GHz, 217 GHz, and 353 GHz. For each frequency, we use the corresponding detector-split maps to form cross-spectra and mitigate noise bias. We adopt the baseline CO and point-source mask of Ref.~\cite{Eskilt:2022:biref-const} that covers approximately $92\%$ of the sky.

Following the methodology of Ref.~\cite{Eskilt:2022:biref-const}, we compute the observed $EB$ power spectra using the {\tt Polspice} package \cite{Chon:2003:Polspice}. The spectra are calculated by cross-correlating different detector maps over the multipole range $51 \leq \l \leq 1490$, using the same sky mask as in Ref.~\cite{Eskilt:2022:biref-const}. We apply corrections for the instrumental beam and pixel window function via deconvolution. The resulting power spectra are then binned into evenly spaced multipole bins across the full range for subsequent analysis.

\subsection{Likelihood}

We follow the likelihood approach developed in Refs.~\cite{Minami:2020:method,Diego-Palazuelos:2022,Eskilt:2022:biref-const} to constrain ALP model parameters. Specifically, we use all cross-spectra between different map pairs, excluding auto-correlations to avoid noise bias.

At each multipole $\l$, we define the following residual vector:
\al{
    \{\vec{v}_\l\}_\alpha &\equiv \vec{A}_{\l,ij}^T\begin{pmatrix}
        \hat{C}_\l^{E_iE_j} \\ \hat{C}_\l^{B_iB_j} \\ \hat{C}_\l^{E_iB_j} 
    \end{pmatrix} - \vec{B}_{\l,ij}^T\begin{pmatrix}
        C^{EE}_\l \\ C^{BB}_\l \\ C^{EB}_\l 
    \end{pmatrix}
    \,, \label{Eq:def:v}
}
where $\alpha=(i,j)$ runs over all map pairs with $i\not=j$. 
We then constrain the model parameters $\vec{p}$ by minimizing the residuals through the log-likelihood function \cite{Eskilt:2022:biref-const}: 
\al{
    -2\ln\mC{L}(\vec{p}) = \sum_b\left(\vec{v}_b^T\bR{M}_b^{-1}\vec{v}_b+\ln|\bR{M}_b|\right) 
    \,, 
}
where the sum is over multipole bins $b$, and the covariance matrix $\bR{M}_{b}$ is given by \cite{Eskilt:2022:biref-const}
\begin{widetext}
\al{
    \{\bR{M}_b\}_{\alpha\alpha'} &\equiv \vec{A}_{b,ij}^T \begin{pmatrix}
        {\rm Cov}(\hat{C}_b^{E_iE_j},\hat{C}_b^{E_{i'}E_{j'}}) & {\rm Cov}(\hat{C}_b^{E_iE_j},\hat{C}_b^{B_{i'}B_{j'}}) & {\rm Cov}(\hat{C}_b^{E_iE_j},\hat{C}_b^{E_{i'}B_{j'}}) \\
        {\rm Cov}(\hat{C}_b^{B_iB_j},\hat{C}_b^{E_{i'}E_{j'}}) & {\rm Cov}(\hat{C}_b^{B_iB_j},\hat{C}_b^{B_{i'}B_{j'}}) & {\rm Cov}(\hat{C}_b^{B_iB_j},\hat{C}_b^{E_{i'}B_{j'}}) \\
        {\rm Cov}(\hat{C}_b^{E_iB_j},\hat{C}_b^{E_{i'}E_{j'}}) & {\rm Cov}(\hat{C}_b^{E_iB_j},\hat{C}_b^{B_{i'}B_{j'}}) & {\rm Cov}(\hat{C}_b^{E_iB_j},\hat{C}_b^{E_{i'}B_{j'}})
    \end{pmatrix} \vec{A}_{b,i'j'}
    \,. \label{Eq:def:covariance}
}    
\end{widetext}
The covariance for the binned power spectra is calculated as \cite{Eskilt:2022:biref-const}
\al{
    {\rm Cov}(\hat{C}_b^{XY},\hat{C}_b^{ZW}) = \frac{1}{\Delta \l^2}\sum_{\l\in b}{\rm Cov}(\hat{C}_\l^{XY},\hat{C}_\l^{ZW}) 
    \,, 
}
where the bin size is $\Delta\l=20$ and the unbinned covariance at multipole $\l$ is given by
\cite{Eskilt:2022:biref-const}
\al{
    {\rm Cov}(\hat{C}_\l^{XY},\hat{C}_\l^{ZW}) = \frac{\hat{C}_\l^{XZ}\hat{C}_\l^{YW}+\hat{C}_\l^{XW}\hat{C}_\l^{YZ}}{(2\l+1)f_{\rm sky}} 
    \,. \label{Eq:cov:cls}
}
Here, $f_{\rm sky}$ denotes the effective sky fraction, computed using Eq.~(22) of Ref.~\cite{Eskilt:2022:biref-const}. Note that we omit the observed $EB$ power spectrum from the right-hand side of \eq{Eq:cov:cls} to avoid large fluctuations that may bias the covariance estimation.
Note that we use a nearly full-sky polarization map and the mode coupling between multipoles is very small. More importantly, we apply the multipole binning that further suppresses residual correlations between multipoles. Since the multipole binning is the same as that used in Ref.~\cite{P16:rot} and is determined so that the mode coupling becomes negligible, we do not include the correlation between different multipole bins.

\subsection{Model parameters}

The $EB$ power spectrum induced by cosmic birefringence depends primarily on two ALP parameters: the logarithmic ALP mass $\mu_\phi\equiv \log_{10} m_{\phi} [\mathrm{eV}]$, and the initial rotation angle $\beta_{\rm ini} = -g\phi_{\rm ini}/2$ \cite{Nakatsuka:2022}. 
In addition to $\mu_{\phi}$ and $\beta_{\rm ini}$, we follow the treatment of instrumental miscalibration and foreground modeling as established in Refs.~\cite{Diego-Palazuelos:2022,Eskilt:2022:biref-const}. Specifically, we include eight miscalibration angles $\alpha_i$, corresponding to the two detector-split maps for each of the four Planck frequency bands (100, 143, 217, and 353 GHz). 
To model the Galactic foreground contribution, we employ a parametric approach with four dust amplitude parameters, $A^{\rm dust}_b$ $(b \in [1,4])$. Each parameter characterizes the dust amplitude $A^{\rm dust}_\l$ from Eq.~\eqref{Eq:dust-EB-model} within a specific range of multipoles: $\l \in [51,130]$, $[131,210]$, $[211,510]$, and $[511,1490]$, respectively.

The theoretical $EB$ power spectrum is computed using the code developed in Refs.~\cite{Nakatsuka:2022,Murai:2022:EDE,Naokawa:2023}, which solves the Boltzmann equations with cosmic birefringence. The code assumes a sufficiently small ALP field amplitude $|\phi_{\rm ini}|$ such that the ALP energy density does not affect the background cosmological evolution. 
To improve computational efficiency, we precompute $C_\l^{EB}$ at $\beta_{\rm ini} = 0.3$\,deg for each fixed value of $\mu_{\phi}$, and obtain spectra for arbitrary $\beta_{\rm ini}$ via rescaling. This procedure is valid in the small-angle approximation, where the power spectrum scales linearly with $\beta_{\rm ini}$, and the resulting constraints on $\mu_\phi$ are independent of the specific reference angle chosen.
To avoid introducing bias, we select a sufficiently dense set of $\mu_\phi$ values for precomputing the $EB$ power spectra, ensuring that the final constraints are insensitive to the specific choice of precomputed mass values.

To explore the posterior distribution of the model parameters, we use the affine-invariant Markov Chain Monte Carlo (MCMC) sampler implemented in the {\tt emcee} package \cite{Foreman-Mackey:2013:emcee}.

\subsection{Priors}

\subsubsection{Mass}

We adopt a flat prior on the logarithmic ALP mass parameter $\mu_\phi$, uniformly distributed over the range $\mu_\phi\in [-29.0, -26.5]$. 
The lower bound of this range is chosen since the behavior of the $EB$ power spectrum is the same for the mass at $\mu_{\phi}\ll -28.0$. For this degeneracy, the posterior with a fixed value of $\mu_\phi$ is identical at $\mu_{\phi}\alt -28.0$. The highly degenerated region $\mu_\phi\ll-28.0$ also leads to the volume effect that leads to a computational difficulty, although we find that the results are almost the same even if we choose the lower limit to $\mu_{\phi}=-30.0$. 
The upper bound is set to limit computational cost, as evaluating the birefringence-induced power spectrum becomes increasingly expensive for higher ALP masses.

\subsubsection{Amplitude}

For the parameter, $\beta_{\rm ini}$, we do not have any specific prior from theory and consider an objective uninformative prior to maximize the constraints from data. One of the common choice in this case is the Jeffreys prior as an objective Bayesian analysis widely used for astronomy (see, e.g., Refs.~\cite{Jeffreys:1961:prior,Heavens:2018:Jeffreys,Diacoumis:2018,Percival:2021:prior}). 

Here, we further approximate the Jeffreys prior so that it can be easily implemented in our calculation. We first summarize the behavior of the $EB$ power spectrum and introduce a rescaled amplitude parameter. For low ALP masses, $\mu_{\phi}\alt-28.0$, the birefringence-induced $EB$ power spectrum closely resembles that from a constant rotation angle across all multipoles considered in our analysis \cite{Sherwin:2021:biref,Nakatsuka:2022}. For higher masses, $\mu_{\phi}\agt-28.0$, for a fixed $\beta_{\rm ini}\simeq 0.3$\,deg, the overall amplitude of the $EB$ power spectrum significantly depends on $\mu_{\phi}$. To account for this variation and maintain a consistent amplitude scale across mass values, we define a suppression factor:
\al{
    F_{\rm sup}(\mu_\phi)\equiv \frac{1}{1440}\sum_{\l=51}^{1490}\frac{|C_\l^{\rm EB}(\mu_\phi)|}{|C_\l^{\rm EB}(\mu_\phi=-33.0)|}
    \,.
}
and normalize the power spectrum accordingly. We then introduce a rescaled amplitude parameter:
\begin{equation}
    A_{\rm EB} \equiv \left( \frac{\beta_{\rm ini}}{0.3\,{\rm deg}} \right) F_{\rm sup}
    \,, \label{Eq:def:AEB}
\end{equation}
which quantifies the magnitude of the $EB$ power spectrum induced by cosmic birefringence.
If we change the amplitude parameter from $\beta_{\rm ini}$ to $A_{\rm EB}$, we find that the Jeffreys prior for $\beta_{\rm ini}$ can be approximated to a flat prior on $A_{\rm EB}$ (see Appendix \ref{app:prior}). Thus, we use $A_{\rm EB}$ and adopt the flat prior for $A_{\rm EB}$.

\subsubsection{Other parameters}

For the remaining nuisance parameters, we follow the treatment in Ref.~\cite{Eskilt:2022:biref-const} and adopt flat-uniform priors on miscalibration angles, $\alpha_i\in[-5\,{\rm deg},5\,{\rm deg}]$, and on the $EB$ dust amplitude, $A^{\rm dust}_b\in[0,1]$.

\section{Results} \label{sec:results}

\begin{figure}[t]
\bc
\includegraphics[width=8.5cm]{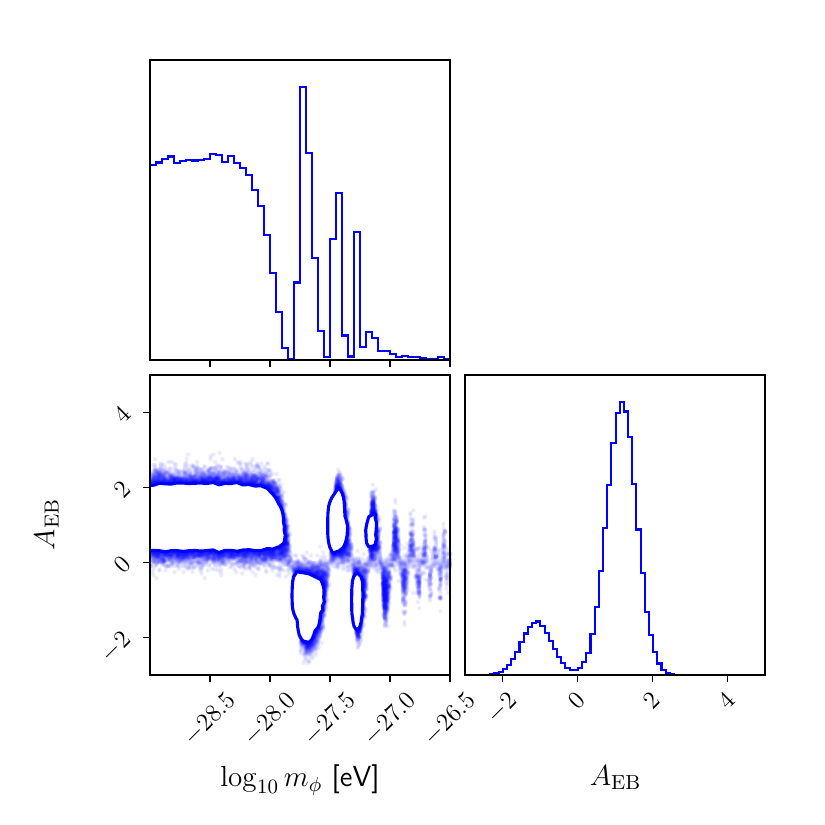}
\caption{
Marginalized posterior distribution of the logarithmic ALP mass $\mu_\phi=\log_{10}m_\phi [\mathrm{eV}]$ and the rescaled amplitude parameter $A_{\rm EB}$, which characterizes the overall strength of the birefringence-induced $EB$ power spectrum. The two-dimensional panel shows the distribution of MCMC samples along with the $2\,\sigma$ contour.
}
\label{fig:const:2params}
\ec
\end{figure}

\begin{figure}[t]
\bc
\includegraphics[width=8.5cm]{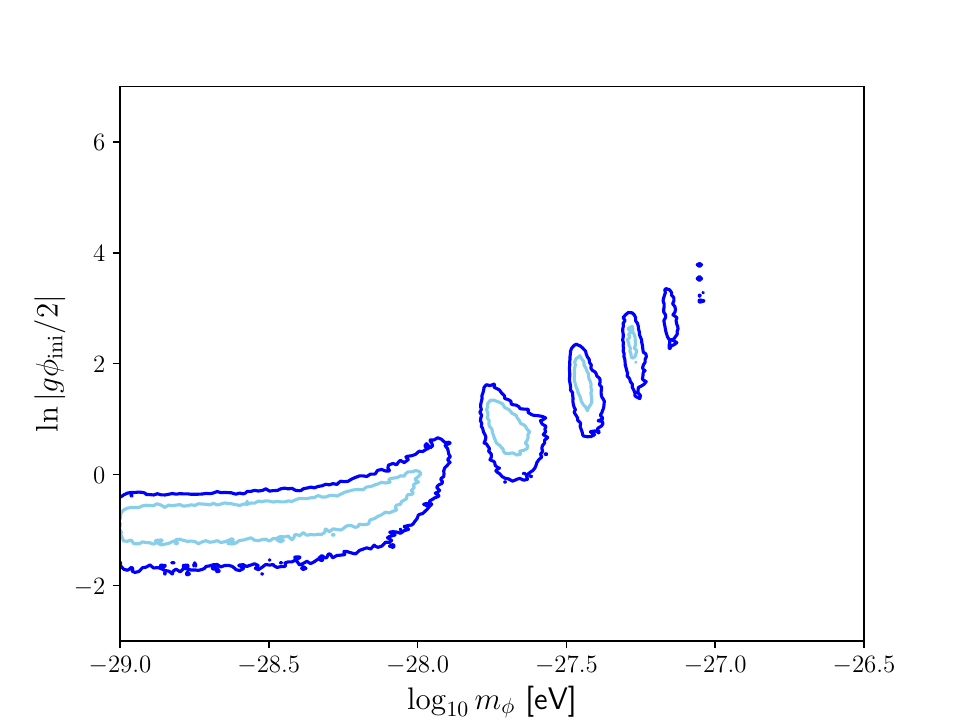}
\caption{
Same as Fig.~\ref{fig:const:2params}, but shown in the $\log_{10}m_\phi[\mathrm{eV}]$–$\ln|g\phi_{\rm ini}/2|$ plane, where $g\phi_{\rm ini}/2$ is expressed in degrees. The two-dimensional posterior is visualized with $1\,\sigma$ (cyan) and $2\,\sigma$ (blue) contours.
}
\label{fig:const:gphi}
\ec
\end{figure}

\begin{figure}[t]
\bc
\includegraphics[width=8.5cm]{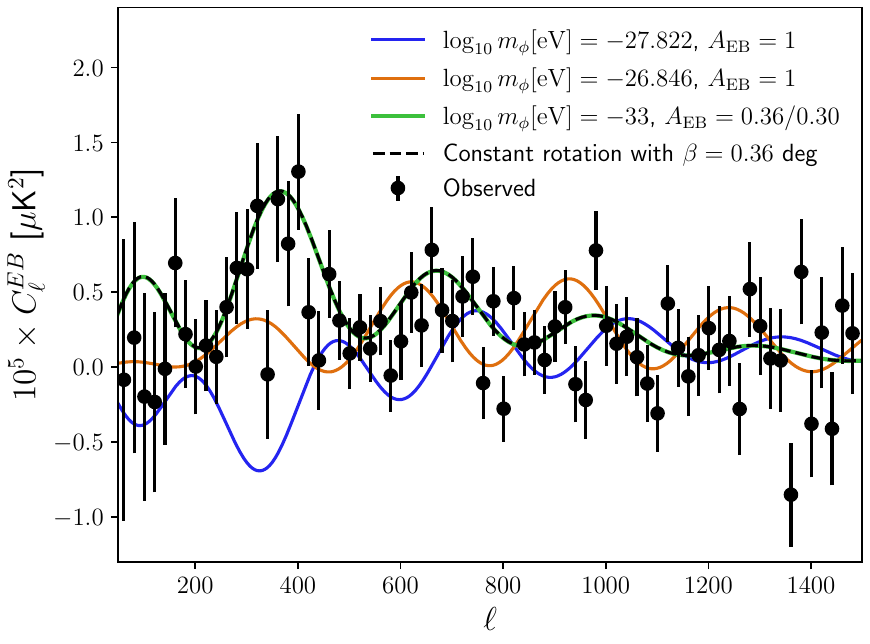}
\caption{
Comparison of theoretical $EB$ power spectra for different ALP masses and amplitudes: $\mu_\phi=-27.822$, $A_{\rm EB}=1$ (blue solid), $\mu_\phi=-26.846$, $A_{\rm EB}=1$ (orange solid), and $\mu_\phi=-33$, $A_{\rm EB}=0.36/0.3$ (green solid). For reference, the spectrum from a constant rotation angle $\beta=0.36\,$deg is shown as a black dashed line. The black points represent the stacked, foreground-subtracted $EB$ power spectrum derived from the data using the best-fit foreground model.
}
\label{fig:EB:highmass}
\ec
\end{figure}

\begin{figure}[t]
\bc
\includegraphics[width=8.5cm]{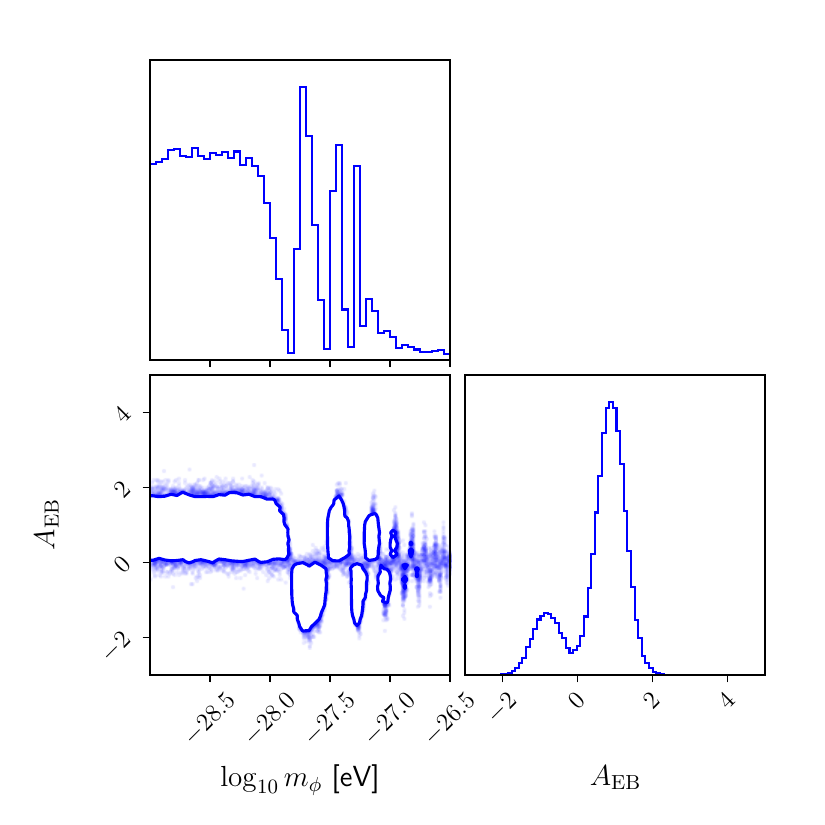}
\caption{
Same as Fig.~\ref{fig:const:2params}, but without modeling the intrinsic dust-induced $EB$ foreground correlation. This comparison illustrates the impact of foreground modeling on the inferred ALP parameters.
}
\label{fig:const:2params:no-dust-model}
\ec
\end{figure}

Figure~\ref{fig:const:2params} presents the parameter constraints in the $\mu_{\phi}$–$A_{\rm EB}$ plane with the $2\,\sigma$ contour obtained from the posterior marginalized over the miscalibration angles and dust amplitude parameters. 
The full posterior distributions for all parameters are provided in Appendix~\ref{app:fullposterior}. 
Figure~\ref{fig:const:gphi} shows the corresponding $1$ and $2\,\sigma$ contours in the $\mu_{\phi}$–$\ln|\beta_{\rm ini}|$ plane where we change the $y$-axis from $A_{\rm EB}$ to $\ln|\beta_{\rm ini}|$ using Eq.~\eqref{Eq:def:AEB}.
We find that the Planck polarization data favor a nonzero isotropic cosmic birefringence induced by ALPs, although the posterior exhibits multiple peaks. Notably, the data exclude the logarithmic ALP masses at $\mu_{\phi} \simeq -27.8$, $-27.5$, $-27.3$, $-27.2$, $-27.1$ with more than $2\sigma$ statistical significance. The mass range $\mu_{\phi}\in [-27.0, -26.5]$ is similarly disfavored at greater than $2\sigma$.
It is important to note that for $\mu_{\phi} \lesssim -28.0$, the $EB$ power spectrum becomes approximately proportional to the $EE$ power spectrum across most multipoles, except near the reionization bump \cite{Sherwin:2021:biref}. Consequently, at $\mu_{\phi} \leq -29.0$, the $EB$ power spectrum retains the same shape across the multipole range accessible to Planck, and this mass range remains consistent with the data.
Regarding the amplitude of the $EB$ power spectrum, $|A_{EB}| \simeq 1$ is favored as expected. In the regions where $A_{EB} \simeq -1$ is favored, $\beta$ has an opposite sign to $\beta_\mathrm{ini}$ around the recombination epoch due to the oscillating behavior of the ALP field. Since the oscillation phase in the recombination epoch shifts depending on the ALP mass, positive and negative $A_\mathrm{EB}$ are alternately favored for $\mu_\phi \gtrsim -28.0$.

To elucidate why the Planck data disfavor higher ALP masses, Fig.~\ref{fig:EB:highmass} presents the $EB$ power spectra for $\mu_{\phi} = -27.822$ and $-26.846$, both assuming $A_{\rm EB} = 1$. For comparison, we also show the power spectrum for $\mu_{\phi} = -33$ with $A_{\rm EB} = 0.36/0.30 = 1.2$, which corresponds to the best-fit value of $A_{\rm EB}$ using the Planck HFI data \cite{Eskilt:2022:biref-const}. Additionally, we include the power spectrum for a constant rotation angle $\beta = 0.36\,\mathrm{deg}$.
The power spectrum for $\mu_{\phi} = -33$ closely matches that of the constant rotation scenario and is in excellent agreement with the observed data. In contrast, the spectra for $\mu_{\phi} = -27.822$ and $-26.846$ exhibit shifts in the acoustic peak structure relative to the $\mu_{\phi} = -33$ case. Notably, at multipoles around $\l\sim 400$, these higher mass cases show significant discrepancies from the observed spectrum.
Such deviations provide a clear basis for excluding these ALP mass values.

We also perform an analysis without explicitly modeling the $EB$ power spectrum from intrinsic dust foregrounds. Specifically, we repeat the same analysis but excluding the dust amplitude parameters $A_b^{\rm dust}$ in the model parameter set and setting $x_\l=A_\l^{\rm dust}\sin4\psi_\l=0$ in \eq{Eq:vecA:final,Eq:vecB:final}. The data prefer a slightly smaller value of $A_{\rm EB}$ than that without the dust $EB$ modeling. The resulting constraints, shown in Fig.~\ref{fig:const:2params:no-dust-model}, are broadly consistent with those obtained when including the dust $EB$ foreground modeling, indicating the robustness of our findings.

\section{Summary and discussion} \label{sec:summary}

We have constrained the mass of ALPs using Planck HFI polarization data, under the assumption that isotropic cosmic birefringence is sourced by ALPs. Our analysis reveals that the data favor mass ranges in which birefringence is effectively described by a constant rotation angle. Consequently, certain mass ranges---specifically $\mu_{\phi}\simeq-27.8$, $-27.5$, $-27.3$, $-27.2$, $-27.1$, as well as $\mu_\phi\in[-27.0,-26.5]$---are excluded at more than $2\sigma$ statistical significance. Importantly, the region $\mu_{\phi}\alt-28.0$ remains unconstrained and allows for the possibility that ALPs play the role of dynamical dark energy, consistent with recent results from DESI \cite{Nakagawa:2025ejs}. We also demonstrated that this conclusion is robust against uncertainties in the modeling of intrinsic $EB$ power spectrum of dust foreground. 

In this work, we assume that ALPs act as spectator fields and do not contribute to the background evolution. For a quadratic potential, Ref.~\cite{Fujita:2020ecn} provides constraints on the ALP-photon coupling constant $g$, based on the requirement that ALPs remain subdominant in energy density. Their analysis uses a benchmark value of $|g\phi_{\rm ini}/2|=0.3\,$deg, ensuring consistency with upper limits on ALP energy density. 
Within the mass range considered in our study, they find a wide range of viable $g$ values consistent with current experimental constraints from CAST, SN1987A, and Chandra. Therefore, our results are compatible with the assumption that the contributions from ALPs to the background evolution are negligible.

Although Ref.~\cite{Fujita:2020ecn} also places constraints on ALP-induced cosmic birefringence, their analysis does not use a full solution of the Boltzmann equations and does not rule out any mass ranges. In contrast, our work provides the first exclusion of specific ALP masses under the assumption that the observed birefringence originates from ALPs, using a full Boltzmann treatment of CMB polarization. 

Our limits on the ALP mass rely on the assumption of the mass potential. If higher-order terms of the potential exist and affect the ALP dynamics, the oscillation phase during the recombination alters, and the constraints on the ALP are modified. While one can carry out an analysis similar to the one presented here, one has to vary an additional parameter in such a case because the degeneracy between $g$ and $\phi_\mathrm{ini}$ is resolved.

We did not investigate whether the $n\pi$-phase ambiguity, recently discussed in Ref.~\cite{Naokawa:2024xhn}, could reconcile the data with ALP masses that are otherwise excluded at more than the $2\,\sigma$ level in the absence of this ambiguity. Accounting for the ambiguity, we can consider the regime where $|\beta(\eta)|\gg 1\,$deg for $\eta$ during the recombination, leading to a breakdown of the small-angle approximation $\sin4\beta(\eta)\simeq 4\beta(\eta)$. 
For the excluded masses, the rotation angle for photons emitted during recombination varies rapidly and significantly, and $EB$ power spectrum has a nontrivial spectral shape. These variations also tend to suppress the polarization signal \cite{Fedderke:2019:biref}, resulting in an $EE$ power spectrum inconsistent with the data, suggesting that such scenarios are unlikely.

Planck polarization data lacks sensitivity to the small angular scales beyond $\l\agt1500$. At these scales, the shape of the $EB$ power spectrum can be further modified by the ALP dynamics during the recombination epoch. Additional high-resolution data, such as that from the Atacama Cosmology Telescope, would provide valuable information at these multipoles and could further tighten constraints on ALP parameters. 

A further low-redshift test of cosmic birefringence is important to uncover the origin of cosmic birefringence. 
For example, cosmic birefringence induced by ALPs with $\mu_\phi \lesssim 10^{-32}\,\mathrm{eV}$ can be probed using the polarization of low-redshift radio galaxies~\cite{Naokawa:2025shr}. Additionally, complementary constraints can be obtained from the polarization and shape of low-redshift galaxies~\cite{Yin:2024:galaxy}. Since this parameter space is challenging to access using CMB data alone, such low-redshift observations provide a valuable and independent avenue for testing cosmic birefringence.


\begin{acknowledgments}
We thank Matthew Johnson, Eiichiro Komatsu, and Blake Sherwin for helpful comments and discussion. 
This work was supported in part by JSPS KAKENHI Grant Numbers JP20H05859 (TN, KM, and FN), JP22K03682 (TN), JP24KK0248 (TN), JP25K00996 (TN), JP23KJ0088 (KM), JP24K17039 (KM), and JP24KJ0668 (FN).
Part of this work uses resources of the National Energy Research Scientific Computing Center (NERSC). The Kavli IPMU is supported by World Premier International Research Center Initiative (WPI Initiative), MEXT, Japan. FN acknowledges the Fore-front Physics and Mathematics Program to Drive Trans-formation (FoPM), a World-leading Innovative Graduate
Study (WINGS) Program, the University of Tokyo.
\end{acknowledgments}


\section*{Data Availability}
The data that support the findings of this article are not publicly available. The data are available from the authors upon reasonable request.

\onecolumngrid

\appendix

\section{Full contours} \label{app:fullposterior}

\begin{figure}[t]
\bc
\includegraphics[width=18cm]{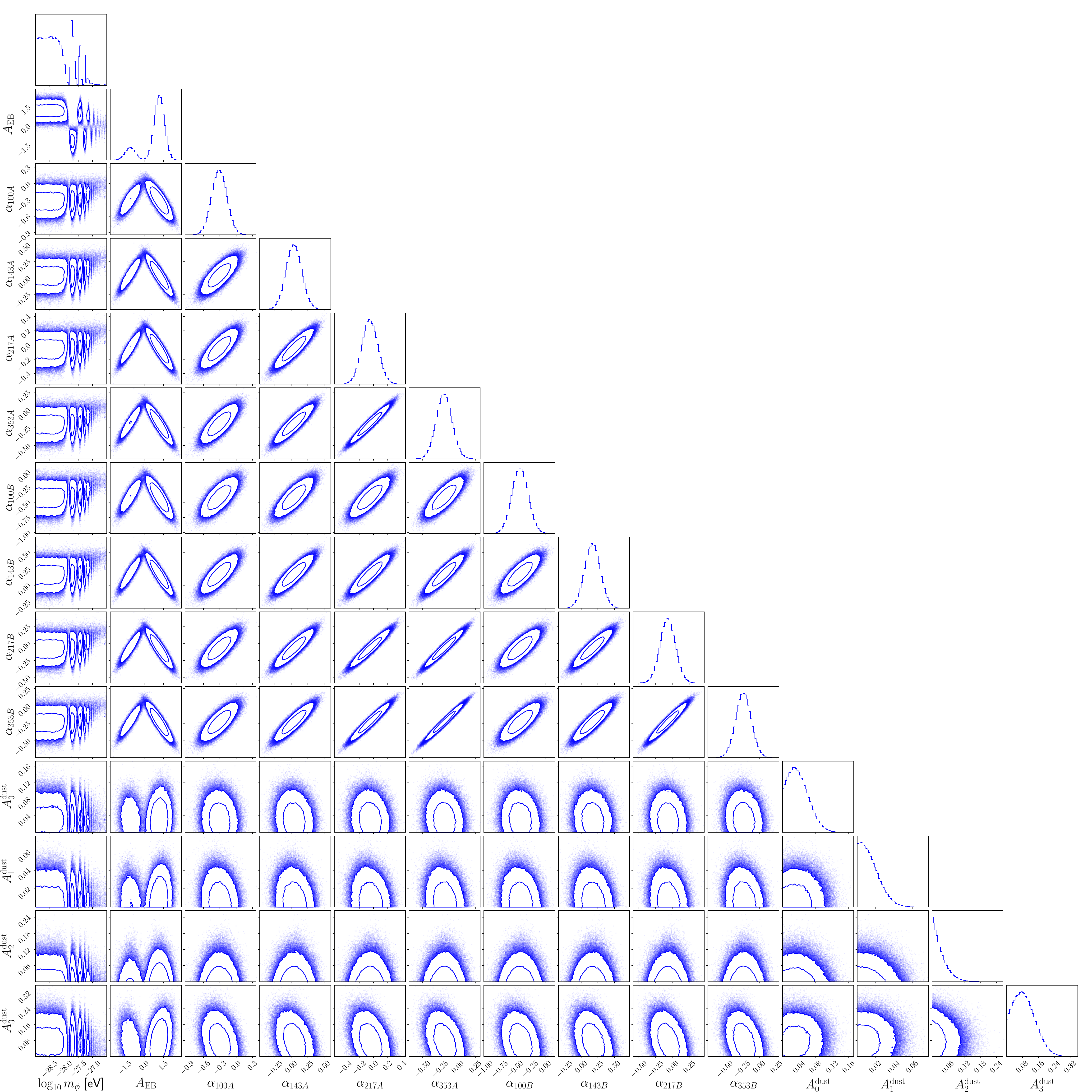}
\caption{
Constraints on the ALP mass, the overall rescaled amplitude of the $EB$ power spectrum, miscalibration angles, $\alpha_i$, and dust $EB$ amplitude, $A^{\rm dust}_b$. 
}
\label{fig:const}
\ec
\end{figure}

Figure~\ref{fig:const} shows the constraints on all parameters in our analysis. 
Most of the miscalibration angles are consistent with zero within $2\,\sigma$ significance, which is similar to the constraints in previous studies \cite{Diego-Palazuelos:2022,Eskilt:2022:biref-const}. 
The constraints on the dust amplitude parameters are close to that obtained in Ref.~\cite{Eskilt:2022:biref-const} where they constrain cosmic birefringence for the constant rotation case.

\section{Derivation of equations} \label{app:eqs}

We here derive the basic equations described in Sec.~\ref{sec:analysis}. Using \eq{Eq:EBij}, we first compute the covariance between observed $E$- and $B$-modes as
\al{
    \bR{C}_{l,ij} &\equiv \begin{pmatrix}
        \hat{C}_\l^{E_iE_j} & \hat{C}_\l^{E_iB_j} \\
        \hat{C}_\l^{B_iE_j} & \hat{C}_\l^{B_iB_j}
    \end{pmatrix}
    = \begin{pmatrix} \hat{E}_{\l m,i} \\ \hat{B}_{\l m,i} \end{pmatrix}
    (\hat{E}^*_{\l m,j},\hat{B}^*_{\l m,j})
    = \bR{R}(\alpha_i)
    \begin{pmatrix}
        C_\l^{EE}+F_\l^{E_iE_j} & C_\l^{EB}+F_\l^{E_iB_j} \\
        C_\l^{EB}+F_\l^{B_iE_j} & C_\l^{BB}+F_\l^{B_iB_j}
    \end{pmatrix}\bR{R}^T(\alpha_j)
    \,. 
}
Here, we assume that the noise in $i$th and $j$th maps are statistically independent and ignore the noise covariance.
Using the formula for the vectorization of the matrix (e.g., Ref.~\cite{Hamimeche:2008ai}), we obtain
\al{
    \begin{pmatrix}
        \hat{C}_\l^{E_iE_j} \\ \hat{C}_\l^{B_iE_j} \\
        \hat{C}_\l^{E_iB_j} \\ \hat{C}_\l^{B_iB_j}
    \end{pmatrix}
    = [\bR{R}(\alpha_j)\otimes\bR{R}(\alpha_i)]
    \begin{pmatrix}
        C_\l^{EE}+F_\l^{E_iE_j} \\ 
        C_\l^{EB}+F_\l^{B_iE_j} \\ 
        C_\l^{EB}+F_\l^{E_iB_j} \\
        C_\l^{BB}+F_\l^{B_iB_j}
    \end{pmatrix}
    \,,
}
where $\otimes$ is the tensor product. Following the previous studies, we exclude the equation for $\hat{C}_\l^{B_iE_j}$ from the above equations and exchange the elements of the vector, yielding
\al{
    \vec{d}_{\l,ij} \equiv \begin{pmatrix}
        \hat{C}_\l^{E_iE_j} \\ \hat{C}_\l^{B_iB_j} \\ \hat{C}_\l^{E_iB_j} 
    \end{pmatrix}
    = \bR{S}_{1,2,3}\bR{P}_{2\leftrightarrow4}[\bR{R}(\alpha_j)\otimes\bR{R}(\alpha_i)]\bR{P}_{2\leftrightarrow4}^T
    \begin{pmatrix}
        C_\l^{EE}+F_\l^{E_iE_j} \\ 
        C_\l^{BB}+F_\l^{B_iB_j} \\
        C_\l^{EB}+F_\l^{E_iB_j} \\
        C_\l^{EB}+F_\l^{B_iE_j}
    \end{pmatrix}
    \,, \label{Eq:app:vecC:0}
}
where $\bR{S}_{1,2,3}$ is the $3\times 4$ selection matrix that selects the first, second, and third elements of a vector, and $\bR{P}_{2\leftrightarrow4}$ is the $4\times 4$ permutation matrix to exchange the second and fourth elements of the vector. The explicit expression of the matrix is given by
\al{
    \tilde{\bR{R}}(\alpha_i,\alpha_j) &\equiv \bR{S}_{1,2,3}\bR{P}_{2\leftrightarrow4}[\bR{R}(\alpha_j)\otimes\bR{R}(\alpha_i)]\bR{P}_{2\leftrightarrow4}^T
    \\ 
    &= \bR{S}_{1,2,3}\bR{P}_{2\leftrightarrow4}\begin{pmatrix}
        \cos2\alpha_i\cos2\alpha_j & -\sin2\alpha_i\cos2\alpha_j & -\cos2\alpha_i\sin2\alpha_j & \sin2\alpha_i\sin2\alpha_j \\
        \sin2\alpha_i\cos2\alpha_j & \cos2\alpha_i\cos2\alpha_j & -\sin2\alpha_i\sin2\alpha_j & -\cos2\alpha_i\sin2\alpha_j \\
        \cos2\alpha_i\sin2\alpha_j & -\sin2\alpha_i\sin2\alpha_j & \cos2\alpha_i\cos2\alpha_j & -\sin2\alpha_i\cos2\alpha_j \\
        \sin2\alpha_i\sin2\alpha_j & \cos2\alpha_i\sin2\alpha_j & \sin2\alpha_i\cos2\alpha_j & \cos2\alpha_i\cos2\alpha_j        
    \end{pmatrix}
    \bR{P}_{2\leftrightarrow4}^T
    \\
    &= \begin{pmatrix}
        \cos2\alpha_i\cos2\alpha_j & \sin2\alpha_i\sin2\alpha_j & -\cos2\alpha_i\sin2\alpha_j & -\sin2\alpha_i\cos2\alpha_j \\
        \sin2\alpha_i\sin2\alpha_j & \cos2\alpha_i\cos2\alpha_j & \sin2\alpha_i\cos2\alpha_j & \cos2\alpha_i\sin2\alpha_j \\
        \cos2\alpha_i\sin2\alpha_j & -\sin2\alpha_i\cos2\alpha_j & \cos2\alpha_i\cos2\alpha_j & -\sin2\alpha_i\sin2\alpha_j
    \end{pmatrix}
    \,.
}
The matrices and vectors in the previous studies \cite{Minami:2020:method,Diego-Palazuelos:2022,Eskilt:2022:biref-const} are given by
\al{
    \bR{R}(\alpha_i,\alpha_j) 
    &\equiv \tilde{\bR{R}}_{1:2,1:2}(\alpha_i,\alpha_j) 
    = \begin{pmatrix}
        \cos2\alpha_i\cos2\alpha_j & \sin2\alpha_i\sin2\alpha_j \\
        \sin2\alpha_i\sin2\alpha_j & \cos2\alpha_i\cos2\alpha_j 
    \end{pmatrix}
    = \frac{1}{2}\begin{pmatrix}
        \cos2\delta_{ij}+\cos2\theta_{ij} & \cos2\delta_{ij}-\cos2\theta_{ij} \\
        \cos2\delta_{ij}-\cos2\theta_{ij} & \cos2\delta_{ij}+\cos2\theta_{ij} 
    \end{pmatrix}
    \,,
    \\ 
    \vec{R}(\alpha_i,\alpha_j) 
    &\equiv [\tilde{\bR{R}}_{3,1:2}(\alpha_i,\alpha_j)]^T
    = \begin{pmatrix}
        \cos2\alpha_i\sin2\alpha_j \\ 
        -\sin2\alpha_i\cos2\alpha_j
    \end{pmatrix}
    = -\frac{1}{2}\begin{pmatrix}
        \sin2\delta_{ij}-\sin2\theta_{ij} \\ 
        \sin2\delta_{ij}+\sin2\theta_{ij}
    \end{pmatrix}
    \,,
    \\
    \bR{D}(\alpha_i,\alpha_j) 
    &\equiv \tilde{\bR{R}}_{3:4,3:4}(\alpha_i,\alpha_j) 
    = \begin{pmatrix}
        -\cos2\alpha_i\sin2\alpha_j & -\sin2\alpha_i\cos2\alpha_j \\
        \sin2\alpha_i\cos2\alpha_j & \cos2\alpha_i\sin2\alpha_j
    \end{pmatrix}
    = \frac{1}{2}\begin{pmatrix}
        \sin2\delta_{ij}-\sin2\theta_{ij} & -\sin2\delta_{ij}-\sin2\theta_{ij} \\
        \sin2\delta_{ij}+\sin2\theta_{ij} & -\sin2\delta_{ij}+\sin2\theta_{ij}
    \end{pmatrix}
    \,,
    \\ 
    \vec{D}(\alpha_i,\alpha_j) 
    &\equiv [\tilde{\bR{R}}_{3,3:4}(\alpha_i,\alpha_j)]^T
    = \begin{pmatrix}
        \cos2\alpha_i\cos2\alpha_j \\ 
        -\sin2\alpha_i\sin2\alpha_j
    \end{pmatrix}    
    = \frac{1}{2}\begin{pmatrix}
        \cos2\delta_{ij}+\cos2\theta_{ij} \\ 
        -\cos2\delta_{ij}+\cos2\theta_{ij}
    \end{pmatrix}
    \,,
}
where $\theta_{ij}=\alpha_i+\alpha_j$ and $\delta_{ij}=\alpha_i-\alpha_j$. 
We decompose \eq{Eq:app:vecC:0} into the blocks that contain the $E$- and $B$-mode auto spectra, and that have $EB$ cross spectra, yielding
\al{
    \vec{\bm{d}}_{\l,ij}
    &= \begin{pmatrix} \bR{R}(\alpha_i,\alpha_j) \\ \vec{R}^T(\alpha_i,\alpha_j) \end{pmatrix}
    \begin{pmatrix}
        C_\l^{EE}+F_\l^{E_iE_j} \\ 
        C_\l^{BB}+F_\l^{B_iB_j}
    \end{pmatrix}
    + \begin{pmatrix} \bR{D}(\alpha_i,\alpha_j) \\ \vec{D}^T(\alpha_i,\alpha_j) \end{pmatrix}
    \begin{pmatrix}
        C_\l^{EB}+F_\l^{E_iB_j} \\
        C_\l^{EB}+F_\l^{B_iE_j}
    \end{pmatrix}
    \,. 
}
The last term in the above equation contains $C_\l^{EB}$ twice. We simplify the last term and find Eq.~\eqref{Eq:vecC:general}. 

Next, we derive Eq.~\eqref{Eq:chisq} from Eq.~\eqref{Eq:vecC:psil}.
We write \eq{Eq:def:Lambda} as 
\al{
    \begin{pmatrix} \bR{\Lambda} \\ \vec{\Lambda}^T \end{pmatrix}
    = 
    \frac{1}{2}
    \begin{pmatrix} 
        \cos2\delta_{ij}+\cos2\theta_{ij} - 2\tan2x_\l\sin2\theta_{ij} & \cos2\delta_{ij}-\cos2\theta_{ij} \\ 
        \cos2\delta_{ij}-\cos2\theta_{ij} + 2\tan2x_\l\sin2\theta_{ij} & \cos2\delta_{ij}+\cos2\theta_{ij} \\ 
        -\sin2\delta_{ij}+\sin2\theta_{ij} + 2\tan2x_\l\cos2\theta_{ij} & -\sin2\delta_{ij}-\sin2\theta_{ij} 
    \end{pmatrix}
    \,. 
}
As \eq{Eq:vecC:psil} has three equations, we eliminate $F_\l^{E_iE_j}$ and $F_\l^{B_iB_j}$ to obtain a single equation:
\al{
    &\hat{C}_\l^{E_iB_j} - \vec{\Lambda}^T\bR{\Lambda}^{-1} \begin{pmatrix}
        \hat{C}_\l^{E_iE_j} \\ \hat{C}_\l^{B_iB_j} 
    \end{pmatrix}
    = [\vec{R}^T - \vec{\Lambda}^T\bR{\Lambda}^{-1} \bR{R}]
    \begin{pmatrix}
        C_\l^{EE} \\
        C_\l^{BB}
    \end{pmatrix}
    + \left[\cos2\theta_{ij}-\vec{\Lambda}^T\bR{\Lambda}^{-1}\begin{pmatrix} -1 \\ 1 \end{pmatrix}\sin2\theta_{ij}\right]C_\l^{EB}
    \,. \label{Eq:app:EE-BB-EB:psil:0} 
}
Using $\tilde{x}_{\l,ij}=\cos2\theta_{ij}-\tan 2x_\l\sin2\theta_{ij}=\cos2\tilde{\theta}_{ij,\l}/\cos2x_\l$, we compute $\bR{\Lambda}^{-1}$ explicitly as
\al{
    \bR{\Lambda}^{-1} 
    &= 2
    \begin{pmatrix}
        \cos2\delta_{ij}+\cos2\theta_{ij} - 2\tan2x_\l\sin2\theta_{ij} & \cos2\delta_{ij}-\cos2\theta_{ij} \\ 
        \cos2\delta_{ij}-\cos2\theta_{ij} + 2\tan2x_\l\sin2\theta_{ij} & \cos2\delta_{ij}+\cos2\theta_{ij} 
    \end{pmatrix}^{-1}
    \notag \\
    &= \frac{1}{2}\frac{1}{\tilde{x}_{\l,ij}\cos2\delta_{ij}}
    \begin{pmatrix}
        \cos2\delta_{ij}+\cos2\theta_{ij} & -\cos2\delta_{ij}+\cos2\theta_{ij} \\
        -\cos2\delta_{ij}+\cos2\theta_{ij} - 2\tan2x_\l\sin2\theta_{ij} & \cos2\delta_{ij}+\cos2\theta_{ij} - 2\tan2x_\l\sin2\theta_{ij}
    \end{pmatrix}
    \,. 
}
We then multiply the vector, $\vec{\Lambda}$, to the above equation, finding a very simple form:
\al{
    [\vec{\Lambda}^T\bR{\Lambda}^{-1}]^T 
    &= \frac{1}{2\tilde{x}_{\l,ij}\cos2\delta_{ij}}
    \begin{pmatrix} 
        \sin2(\theta_{ij}-\delta_{ij})+\tan2x_\l[1+\cos2(\theta_{ij}-\delta_{ij})] \\
        -\sin2(\theta_{ij}+\delta_{ij})+\tan2x_\l[1-\cos2(\theta_{ij}+\delta_{ij})]
    \end{pmatrix}
    \label{Eq:L^TL^-1} \\
    &= \frac{1}{2\tilde{x}_{\l,ij}\cos2\delta_{ij}}
    \begin{pmatrix} 
        \sin4\alpha_j+\tan2x_\l(1+\cos4\alpha_j) \\
        -\sin4\alpha_i+\tan2x_\l(1-\cos4\alpha_i)
    \end{pmatrix} 
    \\ 
    &= \frac{1}{\cos2\delta_{ij}\cos2\tilde{\theta}_{ij,\l}}
    \begin{pmatrix} 
        \cos2\alpha_j\sin2\tilde{\theta}_{j,\l} \\
        -\sin2\alpha_i\cos2\tilde{\theta}_{i,\l}
    \end{pmatrix}
    \,. \label{Eq:appB:LL}
}
For the coefficient of the $E$- and $B$-mode auto power spectra, we compute
\al{
    \bR{\Lambda}^{-1}\bR{R} 
    = (\bR{R}^{-1}\bR{\Lambda})^{-1}
    &= \left[\bR{I}+\tan2x_\l\sin2\theta_{ij}\bR{R}^{-1}\begin{pmatrix}
        -1 & 0 \\ 1 & 0
    \end{pmatrix}\right]^{-1}
    \\ 
    &= \frac{1}{1-\tan2x_\l\tan2\theta_{ij}}\begin{pmatrix}
        1 & 0 \\ -\tan2x_\l\tan2\theta_{ij} & 1-\tan2x_\l\tan2\theta_{ij} 
    \end{pmatrix}
    \,,
}
and find that
\al{
    \vec{R}^T - \vec{\Lambda}^T\bR{\Lambda}^{-1}\bR{R}
    &= \vec{R}^T -[\vec{R}^T+(\tan2x_\l\cos2\theta_{ij},0)]\frac{1}{1-\tan2x_\l\tan2\theta_{ij}}
    \begin{pmatrix}
        1 & 0 \\ -\tan2x_\l\tan2\theta_{ij} & 1-\tan2x_\l\tan2\theta_{ij} 
    \end{pmatrix}
    \\ 
    &= \frac{-\tan2x_\l}{\tilde{x}_{\l,ij}}(1,0) = \frac{-\sin2x_\l}{\cos2\tilde{\theta}_{ij,\l}}(1,0)
    \,. \label{Eq:appB:R-LLR}
}
Alternatively, we can use Eq.~\eqref{Eq:L^TL^-1} to obtain the above equation: 
\al{
    \vec{\Lambda}^T\bR{\Lambda}^{-1}\bR{R} 
    &= \frac{1}{4\tilde{x}_{\l,ij}\cos2\delta_{ij}}
    \begin{pmatrix} 
        \sin2(\theta_{ij}-\delta_{ij})+\tan2x_\l[1+\cos2(\theta_{ij}-\delta_{ij})] \\
        -\sin2(\theta_{ij}+\delta_{ij})+\tan2x_\l[1-\cos2(\theta_{ij}+\delta_{ij})]
    \end{pmatrix}^T
    \begin{pmatrix}
        \cos2\delta_{ij}+\cos2\theta_{ij} & \cos2\delta_{ij}-\cos2\theta_{ij} \\
        \cos2\delta_{ij}-\cos2\theta_{ij} & \cos2\delta_{ij}+\cos2\theta_{ij} 
    \end{pmatrix}
    \notag \\ 
    &= \frac{1}{2}
    \begin{pmatrix} 
        \sin2\theta_{ij}-\sin2\delta_{ij}+\frac{2\tan2x_\l}{\tilde{x}_{\l,ij}} &
        -\sin2\theta_{ij}-\sin2\delta_{ij}
    \end{pmatrix}
    \,. 
}
Finally, for the third term, we use
\al{
    \cos2\theta_{ij}-\vec{\Lambda}^T\bR{\Lambda}^{-1}\begin{pmatrix} -1 \\ 1 \end{pmatrix}\sin2\theta_{ij}
    = \frac{1}{\tilde{x}_{\l,ij}} = \frac{\cos2x_\l}{\cos2\tilde{\theta}_{ij,\l}}
    \,. \label{Eq:appB:EB}
}
Substituting Eqs.~\eqref{Eq:appB:LL}, \eqref{Eq:appB:R-LLR} and \eqref{Eq:appB:EB} into Eq.~\eqref{Eq:app:EE-BB-EB:psil:0}, we obtain Eq.~\eqref{Eq:chisq}. 

We finally derive Eq.~\eqref{Eq:dij:general} from Eq.~\eqref{Eq:vecC:general}. 
From the three equations in Eq.~\eqref{Eq:vecC:general}, by eliminating $F_\l^{E_iE_j}$ and $F_\l^{B_iB_j}$, we obtain
\al{
    \hat{C}_\l^{E_iB_j} - R^T\bR{R}^{-1} \begin{pmatrix}
        \hat{C}_\l^{E_iE_j} \\ \hat{C}_\l^{B_iB_j} 
    \end{pmatrix} 
    &= [D^T - R^T\bR{R}^{-1} \bR{D}]
    \begin{pmatrix}
        F_\l^{E_iB_j} \\
        F_\l^{B_iE_j}
    \end{pmatrix}
    + \left[\cos\theta_{ij}-R^T\bR{R}^{-1}\begin{pmatrix} -1 \\ 1 \end{pmatrix}\sin\theta_{ij}\right]C_\l^{EB}
    \\ 
    &= [D^T - R^T\bR{R}^{-1} \bR{D}]
    \begin{pmatrix}
        F_\l^{E_iB_j} \\
        F_\l^{B_iE_j}
    \end{pmatrix}
    + \frac{1}{\cos\theta_{ij}} C_\l^{EB}
    \,. \label{Eq:appB:general:1}
}
Note that
\al{
    R^T\bR{R}^{-1} = \lim_{x_\l\to 0} \Lambda^T\bR{\Lambda}^{-1} 
    &= \frac{1}{2\cos2\delta_{ij}\cos2\theta_{ij}}
    (\sin4\alpha_j, -\sin4\alpha_i)
    \,, \label{Eq:appB:general:2}
}
and 
\al{
    D^T - R^T\bR{R}^{-1} \bR{D} 
    = \frac{1}{\cos2\delta_{ij}\cos2\theta_{ij}}(\cos2\alpha_i\cos2\alpha_j,\sin2\alpha_i\sin2\alpha_j)
    \,. \label{Eq:appB:general:3}
}
Substituting Eqs.~\eqref{Eq:appB:general:2} and \eqref{Eq:appB:general:3} into Eq.~\eqref{Eq:appB:general:1}, we find Eq.~\eqref{Eq:dij:general}. 

\section{Prior on \texorpdfstring{$A_{\rm EB}$}{AEB}} \label{app:prior}

Here, we explain how the Jeffreys prior on $\beta_{\rm ini}$ leads to the flat prior on $A_{\rm EB}$ used in our analysis. 
The Jeffreys prior on $\beta_{\rm ini}$ is defined as
\al{
    P_{\rm J}(\beta_{\rm ini}) \equiv \sqrt{-\AVE{\PD{^2\ln\cal{L}}{\beta_{\rm ini}^2}}}
    = \sqrt{\frac{1}{2}\AVE{\PD{^2}{\beta_{\rm ini}^2}\sum_b\left(\vec{v}_b^T(\beta_{\rm ini})\bR{M}_b^{-1}\vec{v}_b(\beta_{\rm ini})+\ln|\bR{M}_b|\right)}}
    \,. \label{Eq:def:Pj}
}
From Eq.~\eqref{Eq:def:v}, we obtain
\al{
    \left\{\PD{\vec{v}_b}{\beta_{\rm ini}}\right\}_{ij} 
    = -\sum_{\l\in b}\vec{B}^T_{\l,ij}\begin{pmatrix}
        0 \\ 0 \\ \bar{C}^{EB}_\l
    \end{pmatrix}
    = -\sum_{\l\in b}\frac{\cos2x_\l}{\cos2\tilde{\theta}_{ij,\l}}\bar{C}^{EB}_{\l}
    = -\frac{\cos2x_b}{\cos2\tilde{\theta}_{ij,b}}\bar{C}^{EB}_b
    \,, 
}
where we use the fact that the $EB$ power spectrum scales with $\beta_{\rm ini}$ and define $C^{EB}_{\l}\equiv\beta_{\rm ini}\bar{C}^{EB}_{\l}$. The last equality comes from our assumption of the multipole dependence of the intrinsic $EB$ dust. We also use Eq.~\eqref{Eq:vecB:final} and ignored the $\beta_\mathrm{ini}$ dependence of $C_\l^{EE}$, which is negligibly small. The Jeffreys prior \eqref{Eq:def:Pj} then becomes
\al{
    P_{\rm J}(\beta_{\rm ini}) = \sqrt{\sum_{b\alpha\beta}\left(\frac{\cos2x_b}{\cos2\tilde{\theta}_{\alpha,b}}\bar{C}^{EB}_b\{\bR{M}_b^{-1}\}_{\alpha\beta}\frac{\cos2x_b}{\cos2\tilde{\theta}_{\beta,b}}\bar{C}^{EB}_b\right)}
    \,. \label{Eq:exactPj}
}
For the cosmic birefringence analysis of the Planck data, the small-angle approximation provides almost the same results as that without the approximation \cite{Diego-Palazuelos:2022}. 
We here simplify the calculation with the small-angle approximation. The above equation then becomes 
\al{
    P_{\rm J}(\beta_{\rm ini}) = \sqrt{\sum_b(\bar{C}^{EB}_b)^2\vec{1}^T\bR{M}_b^{-1}\vec{1}}
    \,, 
}
where $\vec{1}$ is a vector whose elements are all unity. 

Next, we see how the above prior depends on $\mu_\phi$ and other parameters. If we change $\mu_\phi$, Ref.~\cite{Nakatsuka:2022} shows that the amplitude of the $EB$ power spectrum significantly changes and the acoustic peaks are shifted. We thus factorize the amplitude change as follows: 
\al{
    |\bar{C}^{EB}_{\l}(\mu_\phi)| 
    \simeq f(\mu_\phi) |\bar{C}^{EB}_{\l}(\mu_\phi=-33)| 
    \,, \label{Eq:EB-approx}
}
where we ignore the shift of the acoustic peaks in the $EB$ power spectrum since the impact of the peak shift would not significantly change the summation of the multipole bins in the Jeffreys prior. Then, the dependence of the prior on $\mu_{\phi}$ is given by
\al{
    P_{\rm J}(\beta_{\rm ini})
    \simeq 
    f(\mu_\phi)\sqrt{ \sum_b[\bar{C}^{EB}_b(\mu_\phi=-33)]^2\vec{1}^T\bR{M}_b^{-1}\vec{1}}
    \,. \label{Eq:appC:Pj-approx}
}
We also note that, in the small-angle limit, Eq.~\eqref{Eq:vecA:final} becomes $\vec{A}^T_{\l,ij}\simeq (0,0,1)$ and the covariance of Eq.~\eqref{Eq:def:covariance} is independent of the parameters: 
\al{
    \{\bR{M}_b\}_{\alpha\alpha'} \simeq {\rm Cov}(\hat{C}_b^{E_iB_j},\hat{C}_b^{E_{i'}B_{j'}}) 
    \,. 
}
Since the quantities inside the square root of Eq.~\eqref{Eq:appC:Pj-approx} are constant and the Jeffreys prior is proportional to $f(\mu_\phi)$, the posterior is described by 
\al{
    P_{\rm post}(\vec{p}) = {\cal L}(\vec{p})f(\mu_\phi)P_{\rm prior}(\vec{q})
    \,, 
}
where $\vec{p}$ is a vector containing $\mu_\phi$, $\beta_{\rm ini}$, miscalibration angles, and $EB$ dust amplitude parameters. $\vec{q}$ is the same as $\vec{p}$ but without $\beta_{\rm ini}$, and $P_{\rm prior}$ is the prior on $\vec{q}$. 

As a final step, we check that the above posterior $P_{\rm post}(\vec{p})$ is equivalent to that used in our analysis. 
To see this, we use the fact that $F_{\rm sup}(\mu_{\phi})=f(\mu_\phi)$ and rewrite Eq.~\eqref{Eq:def:AEB} with $f(\mu_{\phi})$,
\al{
    A_{\rm EB} = \left(\frac{\beta_{\rm ini}}{0.3 \,{\rm deg}}\right) f(\mu_\phi)
    \,. 
}
If we change the parameter from $\beta_{\rm ini}$ to $A_{\rm EB}$, the Jacobian of this transformation is proportional to $1/f(\mu_\phi)$. Thus, the posterior in the new parameter space containing $A_{\rm EB}$ is given by 
\al{
    P_{\rm post}(\vec{p'}) = {\cal L}(\vec{p'})P_{\rm prior}(\vec{q})
    \,, 
}
where $\vec{p'}$ is the vector $\vec{p}$ with $\beta_\mathrm{ini}$ replaced by $A_\mathrm{EB}$. This is equivalent to the posterior with the flat prior on $A_{\rm EB}$. 

\begin{figure}[t]
\bc
\includegraphics[width=8.5cm]{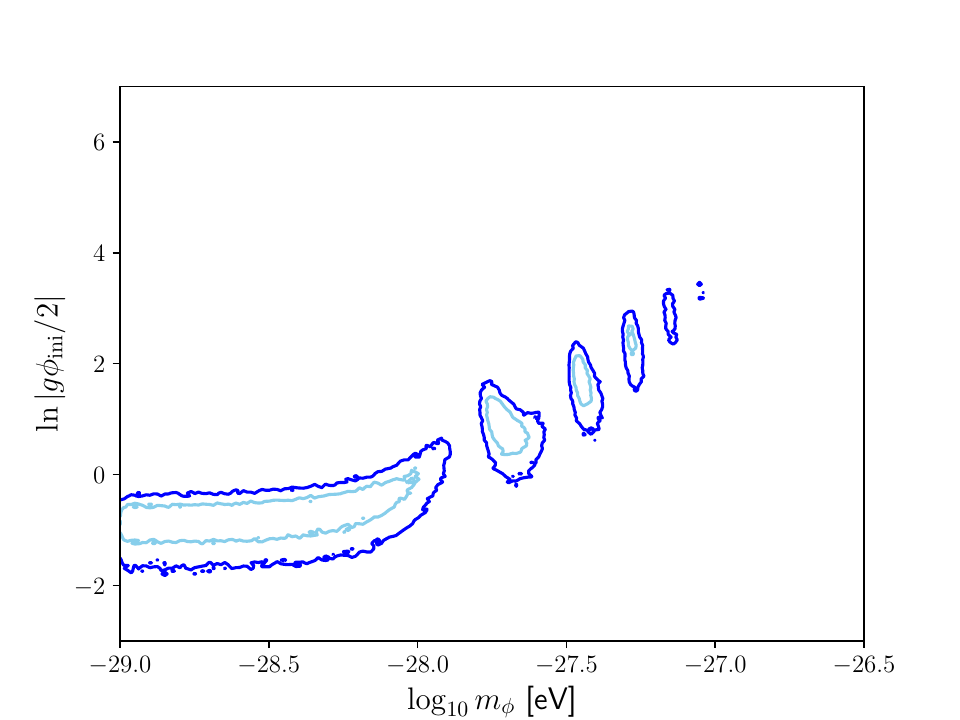}
\caption{
Same as Fig.~\ref{fig:const:gphi}, but with the Jeffreys prior in Eq.~\eqref{Eq:exactPj}. 
}
\label{fig:const:gphi:ePj}
\ec
\end{figure}

To validate the approximation in Eq.~\eqref{Eq:EB-approx}, we further compare our constraint in Fig.~\ref{fig:const:gphi} with that obtained using the Jeffreys prior given in Eq.~\eqref{Eq:exactPj}. The constraint with the Jeffreys prior shown in Fig.~\ref{fig:const:gphi:ePj} is almost identical to that shown in the main text.

\twocolumngrid

\bibliographystyle{mybst}
\bibliography{cite}

\end{document}